\documentstyle[11pt,epsfig]{article}
\title{\bf  From the Schwarzschild Anti de Sitter Black Hole to the Conformal Field Theory}
\author{A. S. Sefiedgar\thanks{e-mail: a.sefiedgar@umz.ac.ir} 
\\ {\small Department of Physics, Faculty of Basic Sciences, University of Mazandaran}\\ {P.O. Box 47416-95447, Babolsar, Iran}}
\begin{document}
\maketitle 
\begin{abstract}
The emergence of the quantum gravitational effects in a very high energy regime necessitates some corrections to the thermodynamics of Black holes. In this letter, we investigate a possible modification to the thermodynamics of Schwarzschild anti de Sitter (SAdS) black holes due to rainbow gravity model. Using the correspondence between a $(d+1)-$dimensional SAdS black hole and a conformal filed theory in $d$-dimensional spacetime, one may find the corrections to the Cardy-Verlinde formula from the modified thermodynamics of the black hole. Furthermore, we show that the corrected Cardy-Verlinde formula can also be derived by redefining the Virasoro operator and the central charge. 
\end{abstract}
\vspace{2cm}
\section{Introduction}
Based on the great discovery of black hole Hawking radiation, it seems that there is a deep connection between three seemingly different branches of science: thermodynamics, gravity and quantum mechanics in the black holes. In fact, Hawking radiation opens an important window to quantum gravity. 
Most of the promising candidates for quantum gravity expect the existence of a minimal observable length at the order of the Planck length \cite{minlength.1,minlength.2,minlength.3,minlength.4,minlength.5a,minlength.5b,minlength.5c}. Therefore, it is natural to take the planck length as a universal constant \cite{unilength}. On the other hand, length is obviously not an invariant under linear Lorenz boost. Therefore, the Lorentz symmetry at Planck scale may not be preserved. The goal of non-linear special relativity or doubly special relativity (DSR) is to preserve the relativity principle and at the same time treat Planck length as an invariant \cite{unilength3}. Within non-linear special relativity, the usual energy-momentum relation may be modified with corrections in the order of Planck length $l_p=\sqrt{8 \pi G}\simeq 1/{M_p}$ as
 \begin{equation}\label{e-m r}
E^2f_1^2(l_pE)-p^2f_2^2(l_pE)=m_0^2,
\end{equation} 
where $f_1$ and $f_2$ are two general functions of energy with a constraint that they approach to unit for the energy scales much less than the Planck scale \cite{unilength4}. The modified dispersion relations may be responsible for threshold anomalies of ultra high energy cosmic rays and gamma ray burst \cite{1-10,1-11,1-12,1-13,1-14,1-15,1-16} and contribute corrections to the black hole thermodynamics \cite{unilength}.
Recently, non-linear special relativity has been generalized to incorporate the effects of gravity, leading to the rainbow gravity model. In rainbow gravity, the metric of the background detected by any probe is not fixed, but depends on the energy of the probe \cite{unilength5}. The rainbow gravity can be used to study the black holes. Of course, the thermodynamical properties of the black hole as well as the final fate of the black hole evaporation may be influenced by the effects of rainbow gravity.

On the other hand, one may relate the black hole thermodynamics with the properties of the conformal field theory (CFT) using the $AdS_{d+1}/CFT_{d}$ and $dS_{d+1}/CFT_{d}$ correspondences. The Cardy-Verlinde (C-V) formula proposed by Verlinde relates the entropy of a certain conformal field theory to its total energy and its Casimir energy \cite{verlinde}. Using $AdS_{d+1}/CFT_{d}$ and $dS_{d+1}/CFT_{d}$ correspondences, this formula holds exactly for different black holes. It means that one may find the quantum gravitational corrections to the C-V formula from the modified thermodynamics of black holes. Previously, the black holes had been considered to modify C-V formula from the generalized uncertainty principle \cite{setare1,setare2}, space non-commutativity \cite{setare3} and modified dispersion relations \cite{sefiedgar1}.

In this paper, we are going to take into account the corrections to the thermodynamics of a $(d+1)-$dimensional Schwarzschild anti de Sitter black hole from the rainbow gravity. Knowing the corrections to the black hole thermodynamics, we can derive the corrections to the C-V formula. It is then shown that the modified C-V formula may also be derived by just redefining the Virasoro operator and the central charge.

\section{A Schwarzschild AdS black hole thermodynamics within rainbow gravity }
The metric of a $(d+1)-$dimensional Schwarzschild anti de Sitter black hole in the rainbow gravity model can be written as
\begin{equation}\label{metric}
ds^2=-\frac{(1-\frac{\mu}{r^{d-2}}+\frac{r^2}{L^2})}{f_1^2}dt^2+\frac{(1-\frac{\mu}{r^{d-2}}+\frac{r^2}{L^2})^{-1}}{f_2^2}dr^2+\frac{r^2}{f_2^2}d\Omega^2_{d-1},
\end{equation}
which is the spherically symmetric solution of $G_{\mu \nu}(E)=8 \pi G(E) T_{\mu\nu}(E)+ \Lambda(E)g_{\mu \nu}(E)$. The newton's constant, $G$, and the cosmological constant, $\Lambda(E)=-3f^2_2(E)/{L^2}$, are energy dependent \cite{9,1}. The parameter $\mu$ is defined as $\mu=\frac{16\pi G_{d+1} M}{(d-1)\Omega_{d-1}}$ and $\Omega_{d-1}=\frac{2\pi^{d/2}}{\Gamma(\frac{d}{2})}$ is the volume of a unit $(d-1)-$sphere.
The position of the horizon, $r_+$, can be derived by solving the equation
\begin{equation}\label{horizon}
\frac{\mu}{r_+^{d-2}}=1+\frac{r_+^2}{L^2}.
\end{equation}
Since investigating the black hole radiation and correction to entropy-area relation is an important subject in theoretical physics, we are going to study the black hole thermodynamics within rainbow gravity.
On the black hole horizon, the surface gravity $\kappa$ can be obtained by 
\begin{equation}\label{T-kappa}
\kappa=-\frac{1}{2} \lim_{r  \rightarrow r_+} \sqrt{-\frac{g^{11}}{g^{00}}}\frac{(g^{00})'}{g^{00}},
\end{equation} 
and it is possible to find the temperature from the surface gravity by $T=\frac{\kappa}{2\pi}$ \cite{1}. 

For a $(d+1)-$dimensional Schwarzschild AdS black hole in the rainbow gravity, one can find the temperature as
\begin{equation}\label{temperature}
T'_{d+1}=\frac{f_2}{f_1} \frac{1}{4\pi}\left[\frac{(d-2)}{r_+}+\frac{dr_+}{L^2}\right], 
\end{equation} 
where $T'$ is the modified temperature via rainbow gravity. It can be seen that the temperature of the modified black hole is different for probes with different energies.
The temperature can be written as 
\begin{equation}\label{TT0}
T'_{d+1}=\frac{f_2}{f_1} T_{d+1}, 
\end{equation}
where
\begin{equation}\label{T}
T_{d+1}=\frac{1}{4\pi}\left[\frac{(d-2)}{r_+}+\frac{dr_+}{L^2}\right],
\end{equation}
is the black hole entropy without any corrections from rainbow gravity and $T'_{d+1}$ is the temperature which is modified by taking into account the effects of $f_1$ and $f_2$. 

Now we are going to define an intrinsic temperature for large modified black holes by identifying probes with radiation particles in the vicinity of the horizon of the black hole. Using the radiation photons with average energy $E=<E>$ and $m_0=0$ to make the measurements, the temperature of the black hole can be identified with the energy of the photons emitting from the black hole as $T \simeq E$ \cite{1}.

To find the black hole temperature, it is necessary to introduce the specific forms of the functions $f_1$ and $f_2$. According to \cite{unilength,1,sefiedgar2}, we can write
\begin{equation}\label{f_1 and f_2}
(f_1)^2=1-\alpha l_p^2E^2, \qquad (f_2)^2=1,
\end{equation}
where $\alpha$ is a positive quantity of order one which have been input to distinguish the correction terms arising from the rainbow gravity effects from the others.  For $\alpha=0$, the energy-momentum relation reduces to its standard form in special relativity.  
By the definitions of $f_1$ and $f_2$ and plugging $T\simeq E$, equation (\ref{TT0}) yields to 
\begin{equation}
{T'}_{d+1}^2=\frac{1}{1-\alpha l_p^2{T'}_{d+1}^2}T_{d+1}^2,
\end{equation}
which can be solved to find the Schwarzschild AdS black hole temperature as
\begin{equation}\label{TTTT}
{T'}_{d+1}=\left[\frac{1-\sqrt{1-4\alpha l_p^2T_{d+1}^2}}{2\alpha l_p^2} \right]^{\frac{1}{2}}.
\end{equation}
For large black holes with $4\alpha l_p^2T_{d+1}^2 \ll 1$, the modified temperature reduces to the temperature of an ordinary Schwarzschild AdS black hole, $T'_{d+1} \simeq T_{d+1}$. For small black holes with extremely high temperature, the temperature reaches its maximal value $T'_{(d+1)max} \simeq \frac{1}{\sqrt{2\alpha}l_p}$ as $T_{d+1} \simeq \frac{1}{2\sqrt{\alpha}l_p}$. Correspondingly the radius of the black hole horizon is bounded from below by $r_+ \geq \frac{\sqrt{\alpha}l_p}{2\pi}$. The existence of a minimum radius leads to the possibility of the existence of the black hole remnant at the late moment of the evaporating process which can be a suitable candidate for dark matter.

By substituting $T_{d+1}$ into equation (\ref{TTTT}), the modified temperature can be written as
\begin{equation}
T'_{d+1}=\frac{1}{\sqrt{2\alpha}l_p} \left[ 1-\sqrt{1-\frac{\alpha l_p^2(d-2)^2}{4 \pi^2 r_+^2}-\frac{\alpha l_p^2d^2r_+^2}{4 \pi^2 L^4}-\frac{\alpha l_p^2d(d-2)}{2 \pi^2 L^2}}  \right]^{\frac{1}{2}}.
\end{equation}
Assuming that the correction terms are small, one may find by some manipulations that 
\begin{equation}\label{inT}
T'_{d+1}=T_{d+1}\sqrt{1+\alpha l_p^2T_{d+1}^2}.
\end{equation}
For large black holes,  the first law of thermodynamics accompanying with equation (\ref{TTTT}) yield to 
\begin{equation}\label{dentropy}
dS'_{d+1}=\frac{dA_{d+1}}{4G_{d+1}}\left[ 1-\alpha l_p^2 \frac{d(d-2)}{16\pi^2 L^2}-\alpha l_p^2 \frac{(d-2)^2}{32\pi^2}\left(\frac{\Omega_{d-1}}{A_{d+1}}\right)^{\frac{2}{d-1}}-\alpha l_p^2 \frac{d^2}{32\pi^2 L^4}\left(\frac{A_{d+1}}{\Omega_{d-1}}\right)^{\frac{2}{d-1}} \right],
\end{equation}
which can be integrated to find $S'_{d+1}$ as the modified entropy of a $(d+1)-$dimensional black hole within rainbow gravity. In equation (\ref{dentropy}), we have considered only the correction terms containing the first power of $\alpha$, without any loss of generality. 

In the case of $d=3$, the entropy of a $(3+1)-$dimensional schwarzschild AdS black hole can be written as
\begin{equation}\label{S4}
S'_4=\frac{A_4}{4G_4}-\frac{3\alpha l_p^2}{64  \pi^2 G_4 L^2} A_4- \frac{\alpha l_p^2}{32\pi G_4} \ln {\frac{A_4}{4G_4}}-\frac{9 \alpha l_p^2}{1024 \pi^3 G_4 L^4}A_4^2,
\end{equation}
where $A_4=\Omega_2r_+^2=4\pi r_+^2$. It is clear from equation (\ref{horizon}) that $r_+$ can be obtained from $\frac{\mu}{r_+}=1+\frac{r_+^2}{L^2}$. 
Furthermore, equation (\ref{inT}) can be expanded to find the temperature for a large $(3+1)-$dimensional schwarzschild AdS black hole as
\begin{equation}\label{T4}
T'_4=\frac{1}{4\pi}\left[\frac{1}{r_+}+\frac{3r_+}{L^2}\right]+\frac{\alpha l_p^2}{2}\left(\frac{1}{4\pi}\right)^3\left[\frac{1}{r_+}+\frac{3r_+}{L^2}\right]^3.
\end{equation}
It is possible to write entropy in an alternative form as
\begin{equation}\label{DeltaS4}
S'_4=S_{4}+\Delta S_4, 
\end{equation}
where
\begin{equation}\label{DeltaS4'}
S_{4}=\frac{A_4}{4G_4}, \qquad   \Delta S_4=-\frac{3\alpha l_p^2}{64  \pi^2 G_4 L^2} A_4- \frac{\alpha l_p^2}{32\pi G_4} \ln {\frac{A_4}{4G_4}}-\frac{9 \alpha l_p^2}{1024 \pi^3 G_4 L^4}A_4^2.
\end{equation}
$S_4$ is the standard entropy in the absence of rainbow gravity effects and $\Delta S_4$ implies the corrections to entropy via rainbow gravity.
The temperature can also be written as
\begin{equation}\label{DeltaT4}
T'_4=T_{4}+\Delta T_4,
\end{equation}
where
\begin{equation}\label{DeltaT4'}
T_{4}=\frac{1}{4\pi}\left[\frac{1}{r_+}  +\frac{3r_+}{L^2}\right],  \qquad  \Delta T_4=\frac{\alpha l_p^2}{2}\left(\frac{1}{4\pi}\right)^3\left[\frac{1}{r_+}+\frac{3r_+}{L^2}\right]^3.
\end{equation}
$T_{4}$ is the standard temperature in the absence of rainbow gravity effects and $\Delta T_4$ implies the corrections to the temperature via rainbow gravity. 

In the case of $d=4$, the entopy of a $(4+1)-$dimensional Schwarzschild AdS black hole can be written as
\begin{equation}\label{S5}
S'_5=\frac{A_5}{4G_5}-\frac{\alpha l_p^2}{8  \pi^2 G_5 L^2} A_5- \frac{3 \alpha l_p^2}{32\pi^2 G_5} (2\pi^2)^{\frac{2}{3}}A_5^{\frac{1}{3}}-\frac{3 \alpha l_p^2}{40 \pi^2 G_5 L^4 (2\pi^2)^{\frac{2}{3}}}A_5^{\frac{5}{3}},
\end{equation}
where $A_5=\Omega_3r_+^3=2\pi^2 r_+^3$ and the temperature is 
\begin{equation}\label{T5}
T'_5=\frac{1}{4\pi}\left[\frac{2}{r_+}+\frac{4r_+}{L^2}\right]+\frac{\alpha l_p^2}{2}\left(\frac{1}{4\pi}\right)^3\left[\frac{2}{r_+}+\frac{4r_+}{L^2}\right]^3.
\end{equation}
The entropy can be written alternatively as
\begin{equation}\label{DeltaS5}
S'_5=S_{5}+\Delta S_5,
\end{equation}
where
\begin{equation}\label{DeltaS5'}
S_{5}=\frac{A_5}{4G_5}, \qquad   \Delta S_{5}=-\frac{\alpha l_p^2}{8  \pi^2 G_5 L^2} A_5- \frac{3 \alpha l_p^2}{32\pi^2 G_5} (2\pi^2)^{\frac{2}{3}}A_5^{\frac{1}{3}}-\frac{3 \alpha l_p^2}{40 \pi^2 G_5 L^4 (2\pi^2)^{\frac{2}{3}}}A_5^{\frac{5}{3}}.
\end{equation}
The temperature can also be written as
\begin{equation}\label{DeltaT5}
T'_5=T_{5}+\Delta T_5,
\end{equation}
where
\begin{equation}\label{DeltaT5'}
T_{5}=\frac{1}{4\pi}\left[\frac{2}{r_+}+\frac{4r_+}{L^2}\right], \qquad   \Delta T_{5}=\frac{\alpha l_p^2}{2}\left(\frac{1}{4\pi}\right)^3\left[\frac{2}{r_+}+\frac{4r_+}{L^2}\right]^3.
\end{equation}
The quantities with prime refers to the modified ones in rainbow gravity while the quantities without prime don't imply any corrections from rainbow gravity. In the case of $d=4$, it is clear from equation (\ref{horizon}) that the radius of the horizon can be derived by $\frac{\mu}{r_+^{2}}=1+\frac{r_+^2}{L^2}$.  

As an important point, one may conclude that the predicted entropy within our formalism is smaller than that of the standard Bekenstein-Hawking entropy while the predicted temperature within our formalism is greater than that of the standard one.
Furthermore, the entropy and the temperature of the black hole can be derived for any other dimensionality in the same manner. 
By continuing the procedure for higher dimensionality, one can deduce that the appearance of the logarithmic correction term in the entropy-area relation is restricted to even black hole dimensionality. In fact, if one tries to consider the higher order correction terms for different values of $d$, the emergence of a logarithmic term in the entropy relation for even-dimensional black hole (odd $d$'s) will be certain. Using $AdS_{d+1}/CFT_{d}$ and $dS_{d+1}/CFT_d$ correspondences, one can utilize the modified thermodynamics to have a more deep insight into CFT. Of course, insisting on the presence of a logarithmic term will put a constraint on the dimensionality of the black hole and its dual CFT. In particular, in the case of $d=3$, the appearance of logarithmic correction term in the entropy-area relation is consistent with the results obtained in string theory and loop quantum gravity \cite{unilength-26,unilength-27,unilength-28,unilength-29}. 

On the other hand, the modified dispersion relation is not confined only to rainbow gravity. In other words, MDR is a common feature to most of the quantum gravity candidates and, in particular, to study of loop quantum gravity (LQG) and of models based on non-commutative geometry. In fact there has been strong interest in modifications to the energy-momentum dispersion relation \cite{unilength4,sefiedgar1-25,sefiedgar1-26,sefiedgar1-27,sefiedgar1-28,sefiedgar1-29a,sefiedgar1-29b,sefiedgar1-29c}. Therefore, the results obtained via MDR, which can be thought as the model independent results, seem to be of importance.

\section{Cardy-Verlinde formula}
The well-known Cardy formula gives the entropy of a 2-dimensional CFT as
\begin{equation}
S_{CFT}=2\pi \sqrt{\frac{c}{6}\left(L_0-\frac{c}{24}\right)},
\end{equation}
where $L_0=ER$ is the product of energy and radius and the shift of $\frac{c}{24}$ is caused by the Casimir effect \cite{sefiedgar1-45}. After making the appropriate identifications for $L_0$ and $c$, the same Cardy formula is also valid for CFT in an arbitrary $d-$dimensional spacetime as
\begin{equation}\label{entropy CFT}
S_{CFT}=\frac{2\pi R}{d-1}\sqrt{E_c(2E-E_c)},
\end{equation}
which is called Cardy-Verlinde formula \cite{verlinde}. $R$ is the radius of the system, $E$ is the total energy and $E_c$ is the Casimir energy defined as
\begin{equation}\label{Ec}
E_c=dE-(d-1)TS.
\end{equation}
We have computed the modified thermodynamics of a $(d+1)-$dimensional Schwarzschild anti de Sitter black hole described by the C-V formula (\ref{entropy CFT}) from rainbow gravity.

Within rainbow gravity, by using equation  
$$E'_c=dE'-(d-1)T'S',$$ 
one can substitute $E'=E+\Delta E$, $T'=T+\Delta T$ and $S'=S+\Delta S$ into the equation
$$S'_{CFT}=\frac{2\pi R}{d-1}\sqrt{E'_c(2E'-E'_c)},$$
to obtain the modified C-V formula as
\begin{eqnarray}\label{SCFTd}
S'_{CFT}=S_{CFT}\left[1+\frac{-\frac{8\pi^2 d}{(d-2)}\Delta T+\frac{4\pi(d-1)^2}{(d-2)}T\Delta S+\frac{8 \pi (d-1)^2}{(d-2)}S \Delta T-2(d-1)^2 T S \Delta S-2(d-1)^2 S^2 \Delta T}{2T[-\frac{4 \pi ^2 d}{(d-2)}+\frac{4\pi (d-1)^2}{(d-2)}S-(d-1)^2S^2]}   \right].
\end{eqnarray}
It is necessary to point that we have replaced $E$ by $T$ from $E=\frac{2\pi}{(d-2)}T$ \cite{sefiedgar1}. Assuming that the correction terms are small, we have ignored their product and used Taylor expansion to derive equation (\ref{SCFTd}). 
Now we can represent the entropy of the conformal field theory which lives in a $d$-dimensional spacetime by $S_{CFTd}$ and it can be derived using the thermodynamical properties of a $(d+1)-$dimensional black hole.

In the case $d=3$, the modified C-V formula can be written as 
\begin{eqnarray}\label{SCFT4-}
S'_{CFT3}=S_{CFT3}\left[ 1+  \frac{-24 \pi^ 2 \Delta T_4+16 \pi T_4\Delta S_4+32 \pi S_4 \Delta T_4-8T_4S_4 \Delta S_4-8S_4^2\Delta T_4}{2T_4(-12\pi ^2+16\pi S_4-4S_4^2)}  \right],
\end{eqnarray}
where $S_{4}$, $T_{4}$, $\Delta S_4$ and $\Delta T_4$ can be substituted from equations (\ref{DeltaS4'}) and (\ref{DeltaT4'}) in terms of the related $r_+$. The modified entropy of the conformal field theory which lives in a $3$-dimensional spacetime have been represented by $S'_{CFT3}$ while $S_{CFT3}$ is the entropy of the CFT in the absence of the rainbow gravity effects. 

In the case $d=4$, the modified C-V formula will be
\begin{eqnarray}\label{SCFT5}
S'_{CFT4}=S_{CFT4}\left[ 1+\frac{-16 \pi ^2 \Delta T_5+18 \pi T_{5} \Delta S_5+36 \pi S_{5} \Delta T_5-18T_{5}S_{5}\Delta S_5 -18S_{5}^2\Delta T_5}{2T_{5}(-8 \pi ^2+18 \pi S_5-9S_5^2)}  \right],
\end{eqnarray} 
where $S_{5}$, $T_{5}$, $\Delta S_5$ and $\Delta T_5$ can be substituted from equations (\ref{DeltaS5'}) and (\ref{DeltaT5'}) in terms of the related $r_+$. The modified entropy of the conformal field theory which lives in a $4$-dimensional spacetime is represented by $S'_{CFT4}$ while $S_{CFT4}$ is the entropy of the CFT in the absence of the rainbow gravity effects.

In 2-dimensional conformal field theory, when the conformal weight of the ground state is zero, the valid form of C-V formula is
\begin{equation}\label{ScL0}
S=2\pi \sqrt{\frac{cL_0}{6}}, 
\end{equation}
where $c$ is the central charge and $L_0$ is the Virasoro operator \cite{setare2, sefiedgar1-14}. If we use $E_cR=(d-1)\frac{S_c}{2\pi}$ in equation (\ref{entropy CFT}), where $S_c$ is the Casimir entropy and drop $E_c$ in analogy with equation (\ref{ScL0}), we obtain the generalization to equation (\ref{ScL0}) for a $d-$dimensional CFT \cite{setare2, sefiedgar1-14} as
\begin{equation}\label{ScL1}
S=\frac{2\pi}{d-1} \sqrt{\frac{cL_0}{6}}, 
\end{equation}
where
\begin{equation}\label{ScL2}
L_0=RE \qquad  ,  \qquad  \frac{c}{6}=\frac{(d-1)S_c}{\pi}=2E_cR. 
\end{equation}
Now equation (\ref{ScL1}) can be applied to a Schwarzschild anti de Sitter black hole. In fact, The C-V formula is the outcome of a striking resemblance between the thermodynamics of CFTs with asymptotically AdS duals and CFTs in two dimensions \cite{setare2, sefiedgar1-14}. Therefore, it is possible to take into account the corrections to C-V formula by just redefining the Virasoro operator and the central charge, the quantities entering the C-V formula.
Using $E=\frac{2\pi}{(d-2)}T$, the Virasoro operator can be modified to 
\begin{equation}
L'_{0}=R(E+\Delta E)=L_0+\frac{2\pi}{(d-2)}R \Delta T.
\end{equation}
The modified central charge  
\begin{eqnarray}
c'=12\left[d(E+\Delta E)-(d-1)(T+\Delta T)(S+\Delta S)\right]R,
\end{eqnarray}
can be written as
\begin{eqnarray}
c'=c+12R \left[\frac{2\pi d}{(d-2)} \Delta T-(d-1)T \Delta S-(d-1) S \Delta T\right],
\end{eqnarray}
where the second orders of the corrections terms have been neglected and the relation $E=\frac{2\pi}{(d-2)}T$ has been applied.

In the case of $d=3$,
\begin{eqnarray}
L'_{0}=L_{0}+\frac{\pi \alpha l_p^2}{(4 \pi)^3}R \left[ \frac{1}{r_+}+\frac{3r_+}{L^2} \right]^3,
\end{eqnarray}
and
\begin{eqnarray}
c'=c+\Delta c,
\end{eqnarray}
where,
\begin{eqnarray}
&\Delta c=\frac{36\pi \alpha l_p^2 R}{(4 \pi)^3}\left[ \frac{1}{r_+}+\frac{3r_+}{L^2} \right]^3 - \frac{6 R}{\pi}\left[ \frac{1}{r_+}+\frac{3r_+}{L^2} \right] \left[-\frac{3\alpha l_p^2}{64  \pi^2 G_4 L^2} A_4- \frac{\alpha l_p^2}{32\pi G_4} \ln {\frac{A_4}{4G_4}}-\frac{9 \alpha l_p^2}{1024 \pi^3 G_4 L^4}A_4^2 \right]+ \\ \nonumber &-\frac{3 \alpha l_p^2 R A_4}{(4\pi)^3G_4} \left[ \frac{1}{r_+}+\frac{3r_+}{L^2} \right]^3.
\end{eqnarray} 

In the case of $d=4$,
\begin{eqnarray}
L'_{0}=L_{0}+\frac{\pi \alpha l_p^2 R}{2 (4\pi)^3} \left[ \frac{2}{r_+}+\frac{4r_+}{L^2} \right]^3,
\end{eqnarray}
and
\begin{eqnarray}
c'=c+\Delta c,
\end{eqnarray}
where
\begin{eqnarray}
&\Delta c= \frac{24 \pi R \alpha l_p^2}{ (4\pi)^3} \left[ \frac{2}{r_+}+\frac{4r_+}{L^2} \right]^3-\frac{9 R}{\pi} [ \frac{2}{r_+}+\frac{4r_+}{L^2} ]\left[-\frac{\alpha l_p^2}{8  \pi^2 G_5 L^2} A_5- \frac{3 \alpha l_p^2(2\pi^2)^{\frac{2}{3}}}{32\pi^2 G_5} A_5^{\frac{1}{3}}-\frac{3 \alpha l_p^2}{40 \pi^2 G_5 L^4 (2\pi ^2)^{\frac{2}{3}}}A_5^{\frac{5}{3}} \right]\\ \nonumber &- \frac{9R \alpha l_p^2A_5}{2G_5(4\pi)^3} \left[ \frac{2}{r_+}+\frac{4r_+}{L^2} \right]^3.
\end{eqnarray} 
Using equation (\ref{ScL1}), the modified Virasoro operator and the modified central charge can be applied to find the corrections to C-V formula for a $d-$dimensional CFT. It is clear that one can find the corrections to the C-V formula for any other CFT dimensionality in the same manner.

\section{Conclusions}
We have computed the thermodynamical properties of a $(d+1)-$dimensional Schwarzschild anti de Sitter black hole within rainbow gravity. The corrected entropy, temperature and energy of the black hole have been used to modify the C-V formula. Since the C-V formula refers to the entropy of a dual conformal field theory living on the $d-$dimensional boundary space, the modified entropy of the CFT has been obtained. We also stressed the point that the C-V formula is the outcome of a resemblance between the thermodynamics of CFTs with asymptotically AdS duals and CFTs in two dimensions. Then we have derived the corrections to the Virasoro operator and the central charge, the quantities which are entering the C-V formula. We have shown the possibility of taking into account the rainbow gravity corrections to the C-V formula by just redefining the Virasoro operator and the central charge.

\end{document}